\documentclass[aps,pre,showpacs,twocolumn,groupedaddress,superscriptaddress]{revtex4}
\usepackage{graphicx}%
\usepackage{amsmath}
\usepackage{dcolumn}% Align table columns on decimal

\begin{document}

\title{Globally clustered chimera states in delay--coupled populations}

\author{Jane H.~Sheeba}%
\affiliation{Centre for Nonlinear Dynamics, School of Physics,
Bharathidasan University, Tiruchirappalli - 620 024, Tamilnadu, India}

\author{V.~K.~Chandrasekar}%
\affiliation{Centre for Nonlinear Dynamics, School of Physics,
Bharathidasan University, Tiruchirappalli - 620 024, Tamilnadu, India}

\author{M.~Lakshmanan}%
\affiliation{Centre for Nonlinear Dynamics, School of Physics,
Bharathidasan University, Tiruchirappalli - 620 024, Tamilnadu, India}

%\date{\today}

\begin{abstract}
We have identified the existence of globally clustered chimera states in delay coupled oscillator
populations and find that these states can breathe periodically, aperiodically and become
unstable depending upon the value of coupling delay. We also find that the coupling delay induces
frequency suppression in the desynchronized group. We provide numerical evidence and theoretical
explanations for the above results and discuss possible applications of the observed phenomena.
\end{abstract}



\maketitle

The existence of chimera states (states characterized by the separation
of identical oscillator groups into synchronized and desynchronized
subgroups) in coupled oscillator populations came as a surprise in the
study of synchronization phenomenon in complex systems. Since its
discovery \cite{Kuramoto:02,Abrams:04}, various theoretical and
numerical developments have been reported on the stability of chimera
states and their existence in systems with varied structures \cite{Abrams:04,Abrams:08},
including time delay \cite{Sethia:08}. By and large, synchronization in coupled
oscillator systems has been analytically and numerically investigated in a rigorous
manner over the past years \cite{Winfree:67,Strogatz:01}.
Possible routes to global synchronization and methods to control
synhronization have also been proposed
\cite{Sherman:92,Rosenblum:04}.
However, complete understanding of the effects induced by coupling delay
in synchronization of coupled oscillator systems is still an open problem.
The consideration of delayed coupling
is vital for modeling real life systems. For example, in a
network of neuronal populations, there is certainly a significant
delay in propagation of signals. In addition there can also be
synaptic and dendritic delays. Other examples include, finite reaction time of chemicals,
finite transfer time associated with the basic mechanisms that regulate gene transcription and mRNA
translation.

In this paper, we demonstrate that coupling delay can induce globally clustered
chimera (GCC) states in systems having more than one coupled identical oscillator (sub) populations. 
By a GCC state, here we mean a state where the
system, which has more than one (sub) population, splits into two different groups, one synchronized and the other desynchronized, each group comprising of oscillators
from both the populations (note that this is in contrast to the chimera state where one of the populations is synchronized while the
other is desynchronized \cite{Abrams:04}). The system under
study is a system of two populations of identical oscillators coupled through a finite delay, represented
by the following equation of motion
\begin{eqnarray}
\dot{\theta_i}^{(1,2)}&=& \omega - \frac{A}{N}
\sum_{j=1}^{N}
f(\theta_i^{(1,2)}(t)-\theta_j^{(1,2)}(t-\tau_1)) \nonumber \\
&&\mp
\frac{B}{N}\sum_{j=1}^{N}
h(\theta_i^{(1,2)}(t)-\theta_j^{(2,1)}(t-\tau_2)).\label{chim_del01}
\end{eqnarray}

\begin{figure}
\begin{center}
\includegraphics[width=8.5cm]{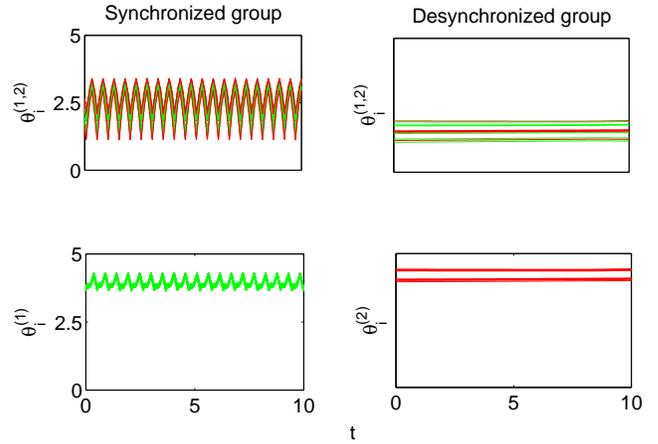}
\caption{(Color online) Occurrence of (stable) GCC in system (\ref{chim_del01}) as explained in the text.
Top panel: Global clustering phenomenon - synchronized and desynchronized (frequency suppressed) groups
have oscillators from both the populations. Bottom panel: One of the populations is synchronized
and the other is desynchronized (frequency suppressed). Green (light gray) and red (dark gray) lines represent oscillators in the first
and the second populations, respectively. Here $\{f,h\}=\{\sin(\theta),\cos(\theta)\}$, $\tau_1=n\tau_2=n\tau$ with $n=1$
(top panel) $A=1.2$, $B=1$ and $\tau=2$, (bottom panel) $A=1.6$, $B=1$ and $\tau=1$.}
\label{Motiv}
\end{center}
\end{figure}

A typical example of such a system is  the two groups of interacting neurons in the brain such as those in the cortex and the
thalamus \cite{Jane:08b}. Here $\omega$ is the natural frequency of the oscillators in the populations and it is the same for
all oscillators in both the populations making all of them identical. However, the two populations are distinguished by
the initial distribution of their phases; the phases are uniformly distributed between $0$ and $\pi$ for the first population and between $\pi$ and
$2\pi$ for the second population. $A$ and $B$ refer to coupling strengths within and between populations,
respectively. The functions $f$ and $h$ are $2\pi$--periodic that
describe the coupling.  $N$ refers to the size of the populations. The complex mean field parameters $X^{(1,2)}+iY^{(1,2)}=r^{(1,2)}e^{i\psi^{(1,2)}}=
\frac{1}{N}\sum_{j=1}^Ne^{i\theta_j^{(1,2)}}$, characterize synchronization within a population
but not global clustering. $\tau_1$ and $\tau_2$ quantify coupling delay within and between populations, respectively.

The investigation is motivated by the numerical discovery of
the existence of GCC states in a system of two identical
populations that are delay--coupled and are given by Eq. (\ref{chim_del01}) (see Fig. \ref{Motiv}). We found that
coupling delay can induce splitting of identical delay--coupled populations
into desynchronized frequency suppressed (vanishing oscillating frequencies) clusters and synchronized clusters.
This splitting can occur either within the populations or between the populations.
The former represents the chimera and the latter is the GCC, as noted earlier. 
Further, the GCC state need not be stable but it can either breathe or can be unstable as will be discussed later.

\begin{figure}
\begin{center}
\includegraphics[width=6cm]{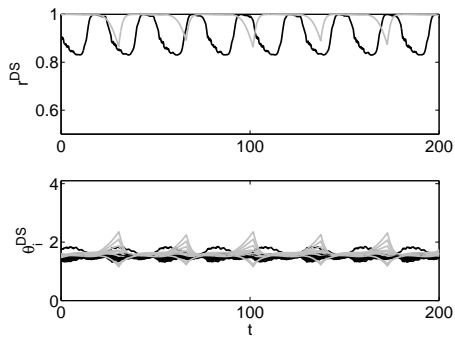}
\caption{Illustration of a breathing GCC state with $A=0.3$, $B=0.2$, $n=1$, $\{f,h\}=\{\sin(\theta),\sin(\theta)\}$ and initial
condition close to the GCC state. Grey and black lines represent the long and
short--periodic breather with $\tau=3.6$ and $\tau=4$, respectively.
Order parameter $r^{\mbox{\tiny{DS}}}$ and the corresponding phases $\theta_i^{\mbox{\tiny{DS}}}$
(see text) are plotted against time in the top and bottom panels, respectively.}
\label{Rplot}
\end{center}
\end{figure}

For illustrative purpose, we simulate system (\ref{chim_del01}) using Runge--Kutta fourth order routine with a time
step of 0.01 (the results are not affected by decreasing the time step below 0.01).
For all the numerical plots shown, we allow a transient time of 2000 units and take $N=32$ (the results have been verified to be independent of the size of the system)
and $\tau_1=n\tau_2=n\tau$, where $n$ is an arbitrary constant. We further found that the GCC
need not be stable but can breathe depending upon the value of the coupling delay. Since the coherence parameter $r$
quantifies synchronization within a population, it can also be used to quantify a breathing or unstable chimera. However, as mentioned earlier, global clustering cannot be quantified using this order parameter. Therefore, in order to quantify a breathing GCC numerically, after allowing the transients, we identify those oscillators whose $\theta_i$s are equal for all times and neglect them so as to end up with the desynchronized group (that comprises oscillators from both the populations, whose size is $N^{\mbox{\tiny{DS}}}$) and calculate its order parameter as
\begin{eqnarray}
\label{dsr}
r^{\mbox{\tiny{DS}}}e^{i\psi^{\mbox{\tiny{DS}}}}=\frac{1}{N^{\mbox{\tiny{DS}}}}\sum_{j=1}^{N^{\mbox{\tiny{DS}}}}
e^{i\theta_j^{\mbox{\tiny{DS}}}},
\end{eqnarray}
where $N^{\mbox{\tiny{DS}}}=2N-N^{\mbox{\tiny{S}}}$. This order parameter $r^{\mbox{\tiny{DS}}}$ can be used to quantify both the chimera and GCC states and also valid for cases where there exists more
than one synchronized cluster. Such multi-cluster states also occur for model (\ref{chim_del01}), the details of which will be published elsewhere. While a GCC is breathing, one of the groups is completely synchronized while the desynchronized group continuously fluctuates. Fig. \ref{Rplot} illustrates breathing GCC where we plot the order parameters $r^{\mbox{\tiny{DS}}}$ for two different values of $\tau$ in the top panel. The grey line represents a long period breather for $\tau=3.6$ where switching occurs between frequency suppressed synchronized state and the desynchronized state. Increasing $\tau$ further to 4 results in a short period breather (the black line) where the desynchronized state oscillates similar to the previous case but in a faster manner.

\begin{figure}
\begin{center}
\includegraphics[width=8cm]{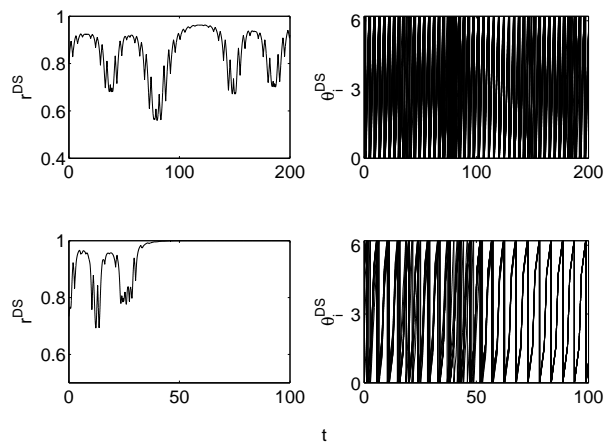}
\caption{Top panel: Aperiodic breathing GCC with $\tau=5$ and $N^{\mbox{\tiny{DS}}}=17$; Bottom panel: Unstable breathing GCC with $\tau=4$ and $N^{\mbox{\tiny{DS}}}=12$, (left) order parameter $r^{\mbox{\tiny{DS}}}$ and (right) the corresponding phases $\theta_i^{\mbox{\tiny{DS}}}$ of the desynchronized group. Here $A=0.6$, $B=0.3$, $n=1$, $N=32$ and $\{f,h\}=\{\sin(\theta),\cos(\theta)\}$.}
\label{LC1}
\end{center}
\end{figure}

\begin{figure}
\begin{center}
\includegraphics[width=6cm]{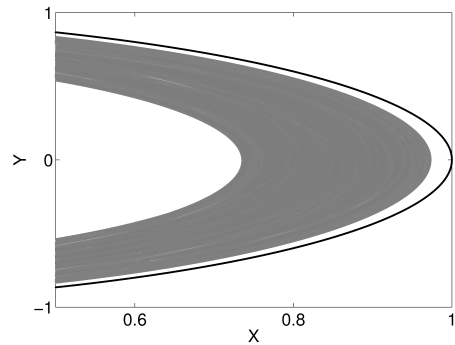}
\caption{Phase portraits showing the limit cycle of the synchronized state (the black line) and a breathing GCC (grey region). The white region between these two represents unstable GCC. The innermost white region represents a stable GCC. Parameter values correspond to Fig. \ref{LC1}. Here $X=r\cos\psi$ and $Y=r\sin\psi$, where $r$ and $\psi$ are the mean-field parameters.}
\label{XY_LC2}
\end{center}
\end{figure}

The GCC can also be unstable where the oscillators in the desynchronized group remain desynchronized for a while after which this state loses its stability and all the oscillators lock to one phase. Thus at this stage the GCC loses stability and a two clustered synchronized state becomes stable. Therefore, finally the system goes from a GCC to a state with two separate synchronized clusters. This phenomenon is depicted in Fig. \ref{LC1}, where for a sufficiently large value of $\tau$ the GCC breathes in an aperiodic manner (top panel). On decreasing $\tau$, this breather loses stability and the desynchronized group entrains itself to a synchronized state (bottom panel). The regions of occurrence of these phenomena in the phase plane, obtained numerically corresponding to Fig. \ref{LC1}, is shown in Fig. \ref{XY_LC2}. The black line is the stable limit cycle attractor of the synchronized group (which is always the same whatever the value of the entrainment frequency of the synchronized group is). The grey region represents the aperiodic breather. The GCC is unstable while in the white region between these two; a GCC in this white region is attracted to the limit cycle and a stable synchronized state is established (as shown in Fig. \ref{LC1} (bottom panel)). A GCC in the innermost white region is always stable. The sizes of all these regions change with respect to system parameters.

\begin{figure}
\begin{center}
\includegraphics[width=6cm]{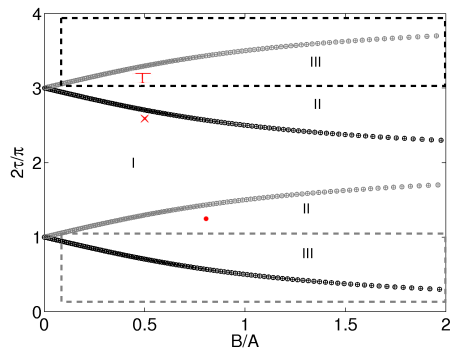}
\caption{(Color online) Stability regions as obtained from theory with $\{f,h\}=\{\sin(\theta),\cos(\theta)\}$. I. Desynchronization, II. Synchronization of the populations individually and III. Global synchronization regions. $\circ$: according to (\ref{chim_del02}), $+$: according to (\ref{chim_del04}) and dotted line: according to (\ref{chim_del05}). The black and grey symbols correspond respectively to the $+$ and $–$ signs in equations (\ref{chim_del02})-(\ref{chim_del05}). Note that the boundaries obtained by the stability analysis on the incoherent ($\circ$) and the synchronization ($+$) regimes are exactly one and the same. The red symbols are the locations of the GCC as from the numerical examples in Figs. \ref{Motiv} and \ref{LC1}. $\bullet$, $\times$ and $\top$ represent the stable, unstable and the breathing GCC respectively.}
\label{Btau1}
\end{center}
\end{figure}

Thus we find that, for a given set of system parameters, increasing/decreasing (depending on the values of the parameters $A$, $B$ and $\tau$, since the behaviour repeats itself periodically as will be discussed later under Fig. \ref{Btau1}) the coupling delay parameter $\tau$ results in the following sequence of GCC dynamics: stable GCC, long--period breather, short--period breather, aperiodic breather and unstable GCC leading to global synchronization. Further increase in $\tau$ from the global synchronization state leads to a stable GCC by following the above mentioned route in the reverse order.

If we are able to discriminate the regions of stability of the synchronized and the
desynchronized states, we will be able to expect the occurrence of GCC near these boundaries with
reference to the numerical observations. This expectation also depends on the fact that the stability of the GCC
state changes periodically with respect to $\tau$ incorporating regions of synchronization and desynchronization.
In order to gain a better understanding of the numerically observed phenomena, we analyze system (\ref{chim_del01}) in the continuum limit $N\rightarrow \infty$. We write down the continuity equation \cite{Winfree:67} for the density of phases $\rho$ and then express $\rho$ and $\{f,h\}$ as Fourier expansions, $\rho=\sum_{k=-\infty}^{\infty}\rho_ke^{ik\theta}$ and  $\{f,h\}=\sum_{k=-\infty}^{\infty}\{f,h\}_ke^{ik\theta}$. Now by considering only the non-trivial $k$th Fourier mode we arrive at the eigen value of that mode $\lambda_k=\bar{A}e^{-\lambda_k\tau_1} \pm \bar{B} e^{-\lambda_k\tau_2} -i\omega_0$ which characterizes the stability of the desynchronized state. Here $\bar{A}=ikf_kA$,
$\bar{B}=ikh_kB$ and $\omega_0=k(\omega-Af_0\mp Bh_0)$. Assuming $\lambda_k=-i\beta/\tau$, one can
find the $kth$ stability region in a parametric form as
%\begin{widetext}
\begin{eqnarray}
\label{chim_del02}
B=\pm kA\frac{|f_k|\cos(n\beta-\alpha_f)}{|h_k|\cos(\beta-\alpha_h)}; \; \;
\tau=\beta/[k(\omega_0 \nonumber \\ +A|f_k|\sin(n\beta-\alpha_f) \pm B|h_k|\sin(\beta-\alpha_h)]^{-1},
\end{eqnarray}
%\end{widetext}
where $\{f,h\}_k=-i|\{f,h\}_k|e^{i\alpha_{\{f,h\}}}$ and $\tau_1=n\tau_2=n\tau$.
The overall stability of the desynchronized state is determined by the overlap of these domains for all the modes.

Now it is also of importance to investigate the stability of the synchronized state for
which we consider the solution to the synchronization state $\theta_i^{(1,2)}=\Omega t$. With this solution,
system (\ref{chim_del01}) becomes $\Omega=\omega-Af(n\Omega \tau)\mp Bh(\Omega \tau)$.
Along with this relation, the condition $Af'(n\Omega \tau)\pm Bh'(\Omega \tau)>0$ should also be satisfied
in order that the synchronized state is stable. This provides the stability regime
\begin{eqnarray}
\label{chim_del04}
B=\frac{\mp Af'(n\beta)}{h'(\beta)}; \; \; \tau=\frac{\beta }{\omega-Af(n\beta)\mp Bh(\beta)},
\end{eqnarray}
where $\beta=\Omega \tau$. The parametric forms (\ref{chim_del02}) and (\ref{chim_del04}) separate the
desynchronization and synchronization regimes.

A homogeneous perturbation $\theta_i^{(1,2)}=
\Omega t+\Delta\theta$ pertaining to the case when all the phases remain equal while their rotation becomes nonuniform in time to the synchronization regimes leads to the following equation for stability
$\Delta\dot{\theta}=-(Af'(n\beta)\pm Bh'(\beta))\Delta\theta +Af'(n\beta)\Delta\theta_{n\tau}\mp Bh'(\beta)\Delta\theta_{\tau}$.
The stability condition for $n=1$ is \cite{DVS:07}
\begin{eqnarray}
\label{chim_del05}
\int_{t_0}^{\infty}[Af'(\beta)\pm Bh'(\beta)-|Af'(\beta)\mp Bh'(\beta)|]dt=\infty.
\end{eqnarray}
The stability of the global synchronization state is determined by the integrand in this condition. From equations (\ref{chim_del02})-(\ref{chim_del05}) it becomes evident that the stability of the synchronized/desynchronized state switches periodically between stable and unstable states depending on the signs of $A$ and $B$, since $h$ and $f$ are $2\pi$ periodic. This is obvious from Fig. \ref{Btau1}, where on increasing $\tau$, regions of synchronization and desynchronization alternate each other.  This is in agreement with the numerical analysis as pointed out earlier and hence forms a theoretical basis. The GCC state can be expected near the stability boundaries shown in Fig. \ref{Btau1}. This is evident from the numerical results depicted in Figs. \ref{Motiv} and \ref{LC1}.

The knowledge about synchrony control
methods is very important because synchronization is desirable
somtimes as in neuronal networks while they support cognition
via temporal coding \cite{Singer:99,Jane:08b} and in the case
of lasers and Josephson junction arrays \cite{Trees:05}.
However, synchronization can also be dangerous in cases like
epileptic seizures \cite{Timmermann:03}, Parkinson's tremor
\cite{Percha:05}, or pedestrians on the Millennium Bridge
\cite{Strogatz:01}. For example, in \cite{Jane:08b} a thalamocortical
model of asymmetrically interacting neuronal populations has
been proposed to simulate the state of emergence from deep to
light an{\ae}sthesia. The model results revealed the fact that
successful coding of information and consciousness is achieved
by the occuurrence of global synchronization between the thalamus
and the cortex. Further, it was eludicated that consciousness and
cognition are kept away during deep an{\ae}sthesia because of the
lack of phase locking between the cortex and the thalamus. This is
one example of a situation where global clustering/synchronization
and hence controlling the same prove to be very crucial. There are
various methods to control synchronization (even its rate and velocity).
However if we could handle it all with one parameter, it makes
life much easier.

In summary, a new type of globally clustered chimera states have been identified in delay coupled populations -- a system of two identical, delay--coupled populations split into two groups, one synchronized and the other desynchronized, each group having a fraction of oscillators from both the populations. We have found that this state need not be stable always but can breathe periodically, aperiodically or become unstable, depending upon the value of coupling delay. A modified version of the order parameter is introduced in order to capture these phenomena. In the presence of coupling delay, frequency suppression is induced in the desynchronized group. We have also provided analytical explanations of the observed effects on the basis of linear stability theory. The illustrative model presented here can be considered as a phenomenological model of oscillatory neural networks. Since coupling delay induces globally clustered chimera, this can prove to be a mechanism for temporal coding of information and cognition and also for memory storage in the nervous system.

%\section*{Acknowledgments}
The work is supported by a Department of Science and Technology (DST), Government of
India -- Ramanna Fellowship program and also by a DST -- IRPHA research project.


\begin{thebibliography}{10}

\bibitem{Kuramoto:02} Y. Kuramoto and D. Battogtokh, Nonlinear Phenom.
Complex Syst. {\bf 5}, 380 (2002); in {\it Nonlinear Dynamics and Chaos: Where
Do We Go From Here?}, edited by S. J. Hogan {\it et al.} (Institute of Physics,
Bristol, England, 2003), p. 209.; S.I.Shima and Y. Kuramoto, Phys. Rev. E. {\bf 69},
036213 (2004).

\bibitem{Abrams:04} D. M. Abrams and S. H. Strogatz, Phys. Rev. Lett.
{\bf 93}, 174102 (2004).

\bibitem{Abrams:08}
D. M. Abrams, R. Mirollo, S. H. Strogatz
and D. A. Wiley, Phys.  Rev. Lett. {\bf 101}, 084103 (2008).

\bibitem{Sethia:08} G. C. Sethia, A. Sen and F. M. Atay, Phys. Rev.
Lett. {\bf 100}, 144102 (2008); O. E. Omel\'chenko, Y. L. Maistrenko and P. A. Tass,
Phys. Rev. Lett. {\bf 100}, 044105 (2008).

\bibitem{Winfree:67} A.~T.~Winfree, {\it J.\ Theor.\ Biol.}
{\bf 16}, 15 (1967); Y. Kuramoto, {\em Chemical Oscillations, Waves, and Turbulence}
  (Springer-Verlag, Berlin, 1984); A. Pikovsky, M. Rosenblum, and J. Kurths, {\em Synchronization -- A
Universal Concept in Nonlinear Sciences} (Cambridge University Press, Cambridge, 2001).

\bibitem{Strogatz:01}
S.~H.~Strogatz, {\it Nature}
{\bf 410}, 268 (2001);

\bibitem{Sherman:92}
A.~Sherman, J.~Rinzel, Proc.\ Natl.\ Acad.\ Sci.\ U.S.A. {\bf 89},
2471 (1992); P. R. Roelfsema, A. K. Engel, P. Konig and W. Singer, Nature {\bf
385}, 6612  (1997); W. Singer, Nature {\bf 397}, 6718 (1999); I.~Z.~Kiss, Y.~M.~Zhai, J.~L.~Hudson,
Science {\bf 296}, 1676 (2002).

\bibitem{Rosenblum:04}
M. G. Rosenblum and A. S. Pikovsky, Phys.\ Rev.\ Lett.\ {\bf 92}, 114102 (2004);
H. Daido, K. Nakanishi, Phys.\ Rev.\ Lett.\ {\bf 96}, 054101 (2006);
Jane~H.~Sheeba, V.~K.~Chandrasekar, A.~Stefanovska, and
P.~V.~E.~McClintock, Phys. Rev. E {\bf 78}, 025201(R) (2008).

\bibitem{Jane:08b}
I. Z. Kiss, M. Quigg, S. H. C. Chun, H. Kori and J. L. Hudson, Biophys. J. {\bf 94}, 1121 (2008);
Jane~H.~Sheeba, A.~Stefanovska, and P.~V.~E.~McClintock,
Biophys. J. {\bf 95}, 2722 (2008).

\bibitem{DVS:07}
D.~V.~Senthilkumar and M.~Lakshmanan, Phys. Rev. E {\bf 76}, 066210 (2007);
M.~Chen and J.~Kurths, Phys. Rev. E {\bf 76}, 036212 (2007).

\bibitem{Singer:99} W.~Singer,  Neuron {\bf 24} 49
(1999); P.~Fries,  Trends.\ Cogn.\ Sci. {\bf 9} 474 (2005);Y.
Yamaguchi, N. Sato, H. Wagatsuma, Z. Wu, C. Molter and Y. Aota,
Curr.\ Opin.\ Neurobiol.  {\bf 17}, 197 (2007).

\bibitem{Trees:05} B. R. Trees, V. Saranathan, D. Stroud, Phys.\ Rev.\
E {\bf 71}, 016215 (2005); F. Rogister, R. Roy, Phys.\ Rev.\ Lett.\
{\bf 98} 104101 (2007).

\bibitem{Timmermann:03}
L. Timmermann, J. Gross, M. Dirks, J. Volkmann, H. Freund and A.
Schnitzler, {\it Brain} {\bf126}, 199 (2003),J. A. Goldberg, T. Boraud, S.
Maraton, S. N. Haber, E. Vaadia, and H. Bergman, {\it J.\ Neurosci.}
{\bf22}, 4639 (2002)

\bibitem{Percha:05}
B. Percha, R. Dzakpasu, M. Zochowski and J. Parent, {\it Phys.\ Rev.\ E.} {\bf 72},
031909 (2005); M. Zucconi, M. Manconi, D. Bizzozero, F. Rundo, C. J.
Stam, L. Ferini-Strambi, R. Ferri, {\it Neurol.\ Sci.} {\bf 26} 199 (2005).

\end{thebibliography}
\end{document}